\documentclass[twocolumn]{aastex61}
\pdfoutput=1 
\usepackage{amsmath,amstext}
\usepackage{mathrsfs}
\usepackage[T1]{fontenc}
\usepackage{apjfonts} 
\usepackage{graphicx}
\usepackage{hyperref}
\usepackage[figure,figure*]{hypcap}
\newcommand{\angstrom}{\mbox{\normalfont\AA}}

\shorttitle{HD~130948BC with ALES}
\shortauthors{Briesemeister et al.}

\begin{document}

\title{High Spatial Resolution Thermal-Infrared Spectroscopy with ALES:\\ Resolved Spectra of the Benchmark Brown Dwarf Binary HD~130948BC}
\correspondingauthor{Zackery W. Briesemeister}
\email{zbriesem@ucsc.edu}

\author{Zackery W. Briesemeister}
\affil{Department of Astronomy and Astrophysics, University of California,
    1156 High St, Santa Cruz, CA 95064 USA}

\author{Andrew J. Skemer}
\affil{Department of Astronomy and Astrophysics, University of California,
    1156 High St, Santa Cruz, CA 95064 USA}

\author{Jordan M. Stone}
\affiliation{Steward Observatory, University of Arizona, 933 N Cherry Ave, Tucson, AZ 85719 USA}

\author{Travis S. Barman}
\affiliation{Lunar and Planetary Laboratory, University of Arizona, 1629 E University Blvd, Tucson, AZ 85721-0092}

\author{Philip Hinz}
\affiliation{Steward Observatory, University of Arizona, 933 N Cherry Ave, Tucson, AZ 85719 USA}

\author{Jarron Leisenring}
\affiliation{Steward Observatory, University of Arizona, 933 N Cherry Ave, Tucson, AZ 85719 USA}

\author{Michael F. Skrutskie}
\affiliation{Department of Astronomy, University of Virginia, 530 McCormick Road, Charlottesville, VA 22904}

\author{Charles E. Woodward}
\affiliation{Minnesota Institute for Astrophysics, University of Minnesota, 116 Church St., S.E. Minneapolis, MN 55455}

\author{Eckhart Spalding}
\affiliation{Steward Observatory, University of Arizona, 933 N Cherry Ave, Tucson, AZ 85719 USA}

\begin{abstract}
We present 2.9--4.1 $\mu$m integral field spectroscopy of the L4+L4 brown dwarf binary HD~130948BC, obtained with the Arizona Lenslets for Exoplanet Spectroscopy (ALES) mode of the Large Binocular Telescope Interferometer (LBTI). The HD~130948 system is a hierarchical triple system, in which the G2V primary is joined by two co-orbiting brown dwarfs. By combining the age of the system with the dynamical masses and luminosities of the substellar companions, we can test evolutionary models of cool brown dwarfs and extra-solar giant planets. Previous near-infrared studies suggest a disagreement between HD~130948BC luminosities and those derived from evolutionary models. We obtained spatially-resolved, low-resolution (R$\sim$20) $L$-band spectra of HD~130948B and C to extend the wavelength coverage into the thermal infrared. Jointly using $JHK$ photometry and ALES $L$-band spectra for HD~130948BC, we derive atmospheric parameters that are consistent with parameters derived from evolutionary models. We leverage the consistency of these atmospheric quantities to favor a younger age ($0.50 \pm 0.07$ Gyr) of the system compared to the older age ($0.79_{-0.15}^{+0.22}$ Gyr) determined with gyrochronology in order to address the luminosity discrepancy.

\end{abstract}


\section{Introduction} \label{sec:intro}

Near-infrared (1--2.5$\mu$m) adaptive optics-fed integral field spectrographs (OSIRIS, \citealt{2006SPIE.6269E..1AL}; GPI, \citealt{2008SPIE.7015E..18M}; SPHERE, \citealt{2008SPIE.7014E..3EC}; Project 1640, \citealt{2011PASP..123...74H}; CHARIS, \citealt{2012SPIE.8446E..9CM}) have been detecting and characterizing exoplanets in high-contrast regimes for nearly a decade (e.g., \citealt{2010ApJ...723..850B, 2011ApJ...733...65B}). Since each wavelength slice of a data cube from an integral field spectrograph (IFS) can be analyzed using techniques for high-contrast image processing, integral field spectrographs are uniquely suited for high-contrast spectroscopy. Furthermore, planet-star spectral diversity can be harnessed to deliver better high-contrast imaging performance and sensitivity to planets compared to more traditional imagers \citep[e.g.,][]{2014A&A...572A..85Z}. 

While IFSs are uniquely capable for obtaining spatially-resolved spectra of exoplanets, adaptive optics-fed IFSs have been confined to the optical and near-infrared ($<3$ $\mu$m). Near-infrared spectra alone are insufficient for precise atmospheric constraints of brown dwarfs and exoplanets due to degeneracies between effective temperature, cloud coverage, convection and non-equilibrium carbon chemistry \citep[e.g.,][]{2009ApJ...702..154S, 2014ApJ...792...17S, 2015ApJ...804...61B}.  Previous works have exploited broad wavelength spectrophotometry extending into the thermal infrared in order to constrain the thermal profiles, compositions, cloud properties and bolometric luminosities of gas-giant planets \citep[e.g.,][]{2011ApJ...729..128C, 2011ApJ...733...65B, 2011ApJ...737...34M, 2012ApJ...754..135M, 2012ApJ...753...14S, 2014ApJ...792...17S, 2014ApJ...794L..15I, 2015ApJ...815..108M, 2017AJ....154...10R}. 

\begin{figure}[h]
\centering
\includegraphics[trim=0cm 1cm 0cm 0cm, clip=False, width=\linewidth]{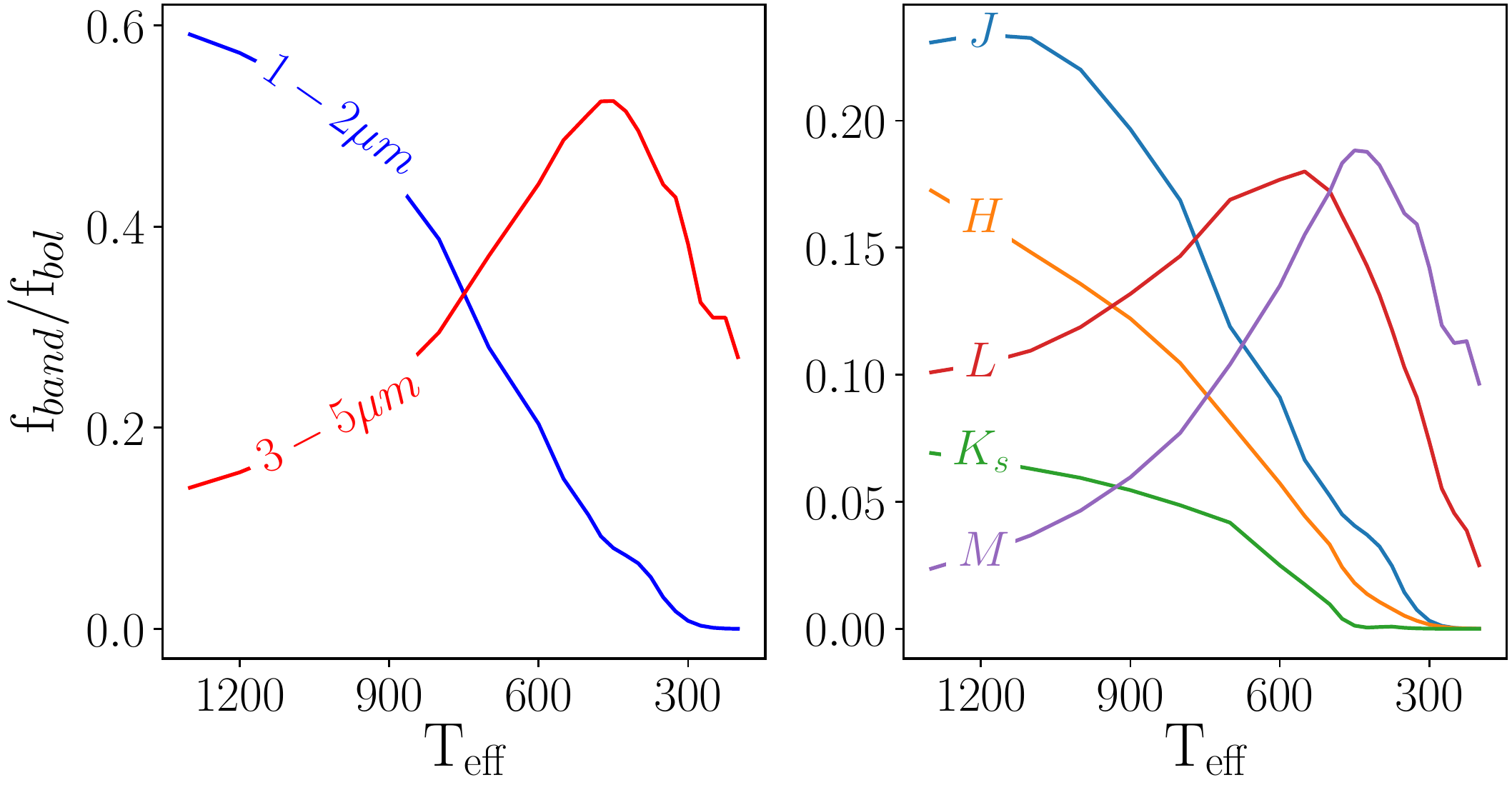} \label{fig:fluxfrac}
\caption{Left: Thermal emission with respect to bolometric flux in 1--2 $\mu m$ and 3--5 $\mu m$ for cold exoplanets. Right: Thermal emission with respect to bolometric flux in each band for cold exoplanets. The 1--2 $\mu m$ contribution to the bolometric flux is negligible for the coldest objects. Models from \citep{2012ApJ...756..172M, 2014ApJ...787...78M}.}
\end{figure}

In the thermal infrared (3--5 $\mu$m), the spectral energy distribution (SED) of gas-giant planets contains a low-opacity atmospheric window that emits a large fraction of a planet's flux \citep{1969BAAS....1..200L, 1986Icar...66..579B, 1997ApJ...491..856B}, especially at cool temperatures (see Figure \ref{fig:fluxfrac}).  Major atmospheric absorbers, such as CH${}_4$, CO and H${}_2$O, have strong absorption features at $\sim$3.3$\mu$m, $\sim$4.7$\mu$m and $\sim$4--5$\mu$m, respectively \citep{2014ApJ...787...78M}. Additionally, the thermal infrared continuum shape is sensitive to cloud thickness and patchiness \citep{2011ApJ...737...34M, 2011ApJ...729..128C, 2014ApJ...792...17S}. 

We built the Arizona Lenslets for Exoplanet Spectroscopy \citep[ALES;][]{2015SPIE.9605E..1DS} to extend the spectroscopic wavelength coverage available for directly imaged planets in order to better understand their atmospheric processes. ALES is the world's first adaptive optics-fed thermal infrared integral field spectrograph (IFS), and exists as a mode of LMIRcam \citep{2010SPIE.7735E..3HS,2012SPIE.8446E..4FL} --- the 1--5 $\mu$m adaptive optics (AO) imager for the Large Binocular Telescope Interferometer \citep[LBTI;][]{2008SPIE.7013E..28H, 2012SPIE.8445E..0UH, 2014SPIE.9146E..0TH}. With ALES, we can exploit these tools developed for high-contrast imaging to probe longer wavelengths and cooler effective temperatures (Figure \ref{fig:fluxfrac}). 

In this work, we present a commissioning data set for ALES: the HD~130948 hierarchical triple system comprised of an L4+L4 brown dwarf binary --- separated by $\lesssim110$~mas --- on a wide orbit ($\sim2\farcs6$) around a sun-like primary star (G2V, [M/H] = 0.0, \citealt{2002ApJ...567L.133P}). The HD~130948 system offers a rare laboratory to test substellar evolutionary models due to the independent measurements of age (from gyrochronology and chromospheric activity of the primary star), luminosity (from photometry and spectroscopy of the brown dwarfs themselves), and total mass of the brown dwarf pair (from orbital motion) \citep{2009ApJ...692..729D}.  While HD~130948B and C are distinct from exoplanets, tests of substellar evolutionary models are key to calibrating models and improving our ability to understand exoplanet observations. Previous comparisons with evolutionary models suggest HD~130948B and HD~130948C are 2 to 3 times more luminous than predicted \citep{2009ApJ...692..729D, 2014ApJ...790..133D, 2017ApJS..231...15D}. In this work, we extend the spatially resolved flux constraints to longer wavelengths, providing a 2.9 to 4.1~$\mu$m spectrum of each component.

In Sections \ref{sec:obs} and \ref{sec:reduction}, we present our ALES observations of HD~130948 along with our reductions.  We extract and calibrate spatially resolved spectra in Section \ref{sec:photometry}.  We fit custom model atmospheres to our spectra, and perform evolutionary model fitting with the derived bolometric luminosities and dynamical mass constraints to determine their individual masses in Section \ref{sec:analysis}. In Section \ref{sec:discussion}, we discuss the implications of our results. We derive bolometric luminosities for each brown dwarf using a combination of near-infrared photometry and ALES $L$-band spectroscopy. The measured bolometric luminosity for each source is brighter than predicted by evolutionary models given the gyro-age of the primary star. We also check the consistency of the evolutionary models with the atmosphere parameters derived in the spectral fitting. We summarize our conclusions in Section \ref{sec:conclusions}. 

\section{Observations} \label{sec:obs}

We observed HD~130948 on 2016 March 26--28 UT as part of early commissioning activities with ALES. We used the left (SX) primary mirror of the two 8.4 meter primary mirrors of the LBT during photometric conditions and sub-arcsecond seeing. Visible light is directed towards the LBTI wavefront sensors for adaptive optics correction \citep{2014SPIE.9148E..03B} performed with the deformable secondary mirror \citep{2011SPIE.8149E..02E}. Wavefront-corrected, diffraction-limited, infrared light is directed into the cryogenic universal beam combiner (UBC) and then into the Nulling and Imaging Camera \citep[NIC;][]{2008SPIE.7013E..39H} where LMIRcam \citep{2010SPIE.7735E..3HS,2012SPIE.8446E..4FL} resides.

For ALES operations, an 8$\times$ Keplerian magnifier, a silicon lenslet array with a pinhole grid to suppress diffraction, a blocking filter and disperser (direct-vision prism assembly) are introduced into the light path via LMIRcam filter wheels. The light incident on the spatial extent of each lenslet is focused through the diffraction-suppressing pinhole grid. Each lenslet sub-image is then dispersed by the direct-vision prism assembly. These dispersed sub-images are imaged onto a 5.2 micron-cutoff Teledyne HAWAII-2RG (H2RG, \citealt{2008SPIE.7021E..0HB}) as a grid of thermal infrared spectra. 

At the time of observation, the FORCAST readout electronics \citep{2010SPIE.7735E..2NL} limited the detector readout to 1024$\times$1024 pixels, instead of the complete 2048$\times$2048 pixels of the H2RG. This subarray contains 50$\times$50 $L$-band spectra with spectral resolution R$\sim$20, covering a field of view of $1\farcs3\times1\farcs3$. \citet{2015SPIE.9605E..1DS} present a description of the design of the version of ALES used in this paper. Subsequent upgrades to ALES, available for current and future use, are described by \citet{2018SPIE10702E..3LH} and \citet{2018arXiv180803301S}.

\begin{table}[h]
\caption{HD~130948 Observations for 2016 March 28}
\centering
\begin{tabular}{c c c c c}
\hline\hline
Observation & $\text{N}_{\text{pointings}}$ & $\text{N}_{\text{frames}}$ & $\text{N}_{\text{coadds}}$ & DIT\footnote{Detector integration time}\\[0.5ex]
\hline
HD~130948A & 24 & 5 & 2 & 1s \\  
HD~130948BC & 24 & 30 & 2 & 1s \\ 
Sky & 24 & 30 & 2 & 1s \\ 
Darks & 24 & 9 & 1 & 1s \\ 
\hline
\end{tabular}
\label{table:obs}
\end{table}

For this dataset, we took natural guide star AO observations of the hierarchical triple system, HD~130948, using HD~130948A as the reference star. The spatial scale of the HD~130948 system is larger than the field of view of ALES (the binary is separated by $2\farcs6$ from the primary). Therefore, the data were obtained in a three point pattern consisting of HD~130948A, sky, and HD~130948BC. Dark frames were obtained between each nod position while the telescope was in motion. A detailed description of this strategy can be found in \citet{2018arXiv180802571S}. Overall, we obtained the data described in Table \ref{table:obs}. Four of the 24 pointings were discarded: two had the binary positioned close to the edge of the lenslet array and two had poor AO correction.

Wavelength calibration of the low-resolution spectra is performed using dome flats at four spectrally unresolved narrowband (R$\sim$100) filters spanning 2.9--3.9 $\mu$m \citep{2018arXiv180802571S}. An empirical dispersion relation is used to propagate the wavelength solution between the four data points for each spectrum.

\section{Reduction} \label{sec:reduction}

\begin{figure*}

\centering
\includegraphics[trim=0cm 0cm 0cm 0cm,clip=False,width=\textwidth]{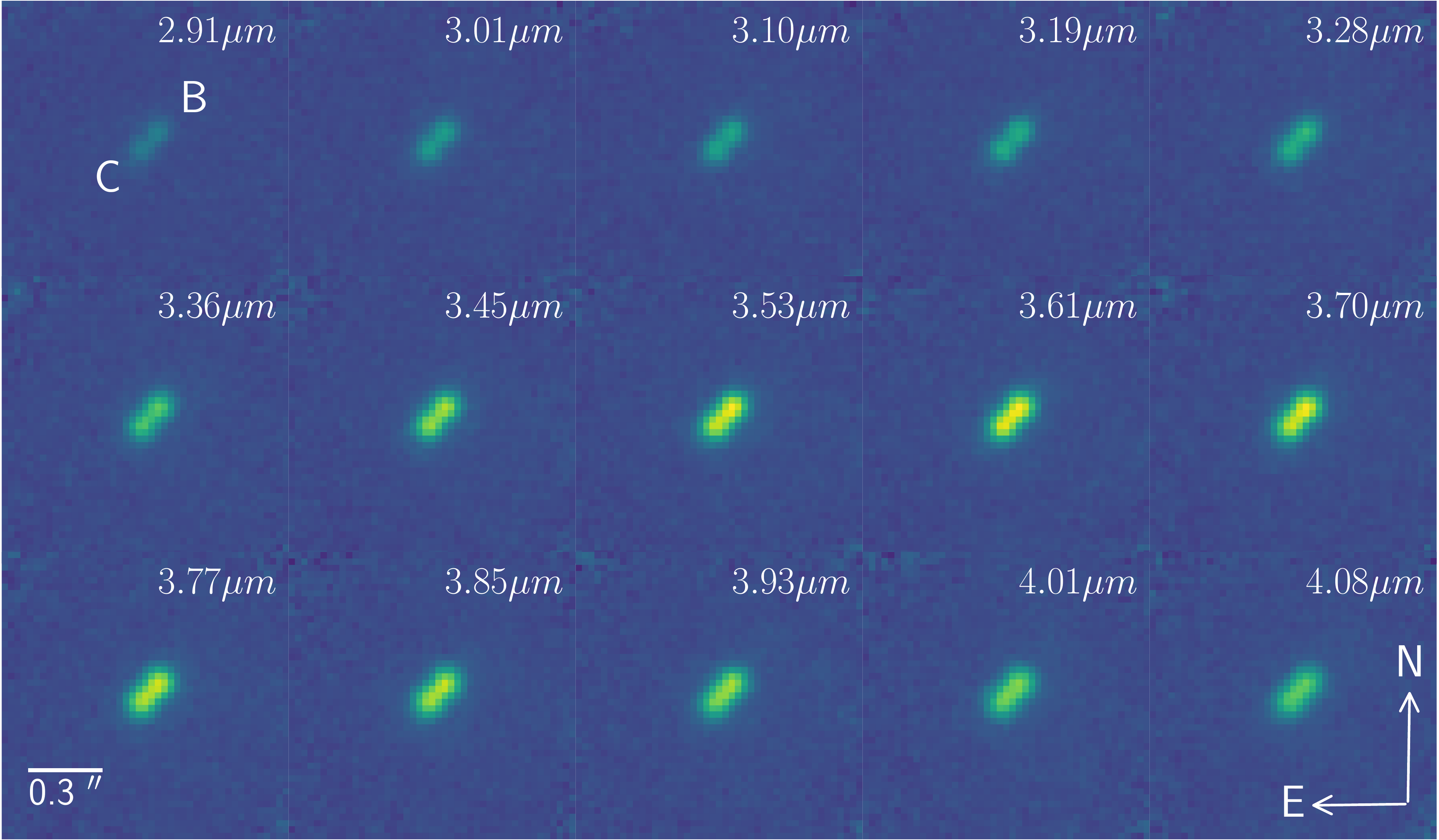}\label{fig:HD130948BCimages}
\caption{ALES data cube of HD~130948BC, where each image is a slice of the cube at the wavelengths specified in the upper right. These are a representative 15 wavelength slices of the 35 wavelength slices in each data cube.}
\end{figure*}

A raw ALES $L$-band frame consists of a grid of 2,500 spectra, each spanning $\sim 37 \times 7$ pixels, dispersed in a grid at a fiducial angle $\theta = \text{tan}^{-1}(\frac{1}{2})$ with respect to detector rows and columns. Raw ALES data were reduced using the ALES Python pipeline (\texttt{MEAD}, \citealt{2018arXiv180803356B}), with the steps performed as follows.

The first step performed by the pipeline is removal of detector artifacts, including the removal of residual detector channel bias, pixel flatfielding, background subtraction, and bad pixel correction. 

For non-linearity correction, we constructed fluence-to-count curves for each pixel by taking sky flats of varying integration times without ALES optics in the light path. The sky flats were bias-subtracted and flatfielded. Outlier pixels were flagged as bad pixels. The linearity correction was applied to all frames. Frames of HD~130948BC had flux corrections $<0.01$\%, sky frames had flux corrections of $<0.5$\% and frames of HD~130948A had flux corrections of $\sim$ 2\%.

The 3.9~$\mu$m narrowband filter calibration data are used to coarsely locate the spectra on the detector. However, the calibration data for observations of HD~130948 on 2016 March 28 were taken at the end of the night; the telescope was set to zenith and the telescope experienced a different gravity vector. The irreproducibility of the multiple filter wheel positions, flexure, and the use of a distinct pupil stop for the calibration data resulted in a field dependent ($\lesssim 1$ pixel) shift between the calibration data and the science data. The deviations of the calibration data from science data, as well as the deviation of dispersion direction from the fiducial angle, are calculated for each spaxel to parameterize a mapping of every light-sensitive pixel to a wavelength calibrated ($\lambda$, y, x) data cube. In order to turn the raw data into wavelength calibrated data cubes, we apply optimal extraction \citep{1986PASP...98..609H} to each spectrum, which becomes the associated spaxel in the data cube.

In a lenslet-based IFS, each lenslet has a slightly different throughput as a function of wavelength. To address this effect, a lenslet flat field is constructed from the normalized, dark-subtracted sky data cube. This is necessary because, by design, LBTI runs in pupil tracking mode only and does not rotate with the sky; astrophysical light of a binary system is incident on different lenslets over the course of observation due to sky rotation.

The cubes containing HD~130948BC and the cubes containing HD~130948A were derotated by the median parallactic angle during each HD~130948BC pointing. Each wavelength slice of every cube was registered with respect to the appropriate wavelength slice of the highest signal-to-noise data cube using the single step Discrete Fourier Transform (DFT) approach outlined in \citep{Guizar-Sicairos2008}. This registration technique calculates pixel-accuracy translations using DFT phase correlation and then upsamples the cross correlation by a factor of 1000 in a 1.5 $\times 1.5$ pixel neighborhood of the estimated pixel shift for subpixel-accuracy translation.

Once the translations of the derotated cubes are determined, the registered, derotated cubes are made with a final interpolation of the original cubes that performs the derotation and registration simultaneously in order to avoid multiple interpolations. The sum of the science and sky variance images undergo the same optimal extraction process to be turned into data cubes, and the respective registration and derotation in a single step. The average of the group of data cubes, weighted by their respective inverse variance cubes, are the final ($\lambda$, y, x) data cubes of HD~130948BC and HD~130948A. The final ($\lambda$, y, x) variance cubes are the mean of all the propagated variance cubes. These final cubes and associated propagated variance cubes are then used for point-spread function (PSF) photometry. 

The final cube for HD~130948BC is shown in Figure \ref{fig:HD130948BCimages}, in which the binary is centered in each wavelength frame denoted by their respective wavelengths. The $\sim110$ mas binary is resolved in $L$-band by ALES. Each image is 50 $\times$ 50 pixels, with one pixel corresponding to 26.1 mas. The binary appears brightest in the middle of the band because the sky transmission of $L$-band peaks at $\sim 3.6 \mu $m. The noise appears worse near the edges of each frame because fewer data cubes overlap in these regions.

\section{PSF Photometry} \label{sec:photometry}

We used PSF fitting to measure the flux from each component at each wavelength slice, explicitly allowing for spatial covariance in the images (See Appendix \ref{sec:fitting} for more details). We took a two-step approach, first fitting for binary separation and position angle, and then extracting flux with the respective priors of those two parameters. A two-step process is required because the position angle $\theta$ and separation $\rho$ of the two brown dwarfs are expected to be constant across all wavelengths, while the position of the binary in the data cube can shift as a function of wavelength due to chromatic optics and atmospheric dispersion. The first step performs PSF subtraction at each wavelength slice completely independently. The second step repeats PSF fitting at each wavelength slice, including Gaussian priors for position angle $\theta$ and separation $\rho$, with hyperparameters derived from the first step.

\begin{figure}
\centering
\includegraphics[trim=0cm 1cm 0cm 0cm, clip=False,width=\linewidth]{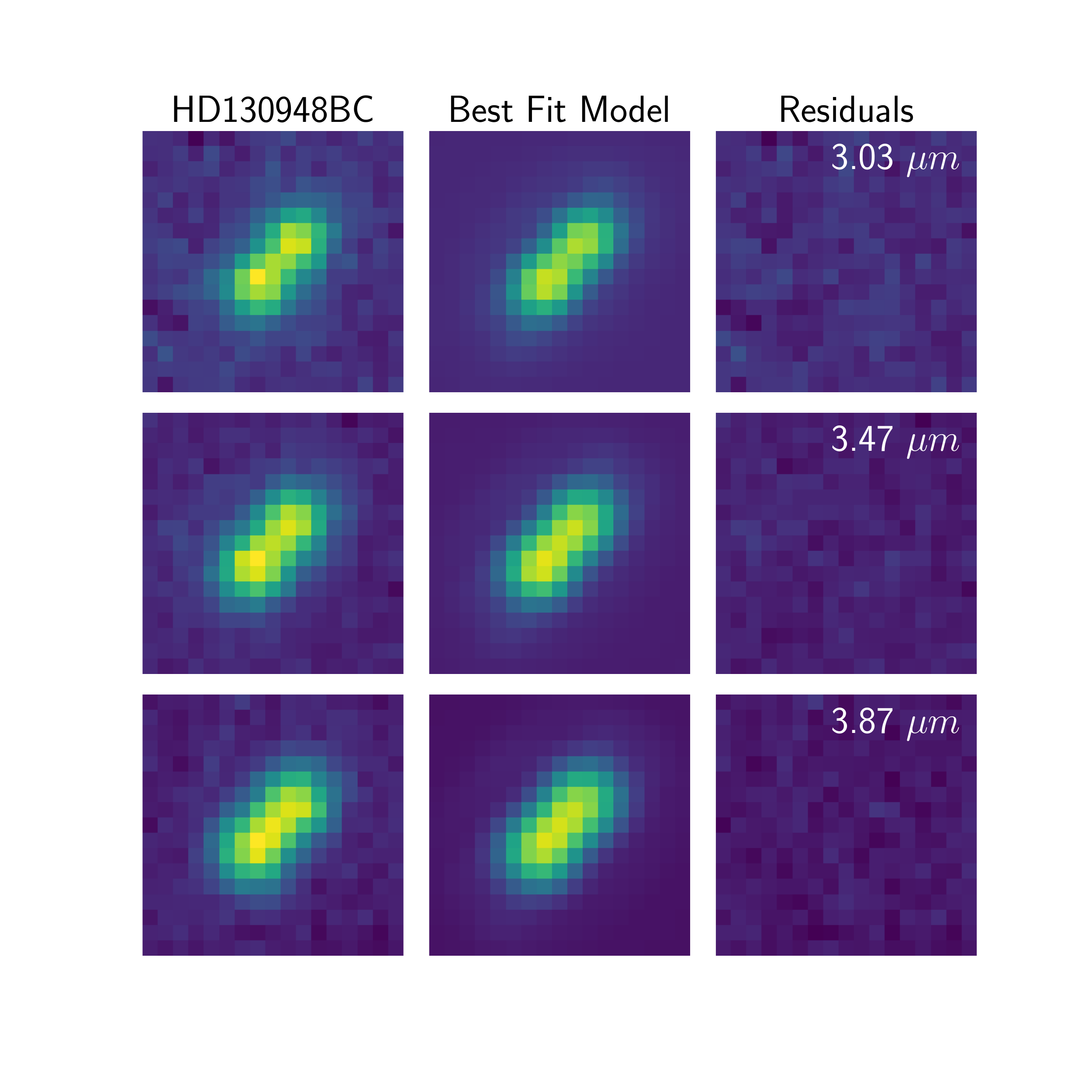} \label{fig:bestfit}
\caption{The first image in each row is a representative data cube slice of HD~130948BC at the wavelength denoted in the upper right corner of the row. The middle image in each row is the best fit model of two scaled PSFs that fit the binary. The right image depicts their residuals. Each wavelength of the cube produced quantitatively similar results. The color stretch is linear.}
\end{figure}

Due to the adaptive optics correction, the HD~130948BC pair is far enough away from the primary star such that high-contrast post-processing algorithms for PSF subtraction are not used, so PSF subtraction does not dominate the correlation of noise in our data. However, interpolation and diffraction still induce spatial and spectral correlations for the HD~130948BC dataset. We extended the Bayesian framework used in \citet{2016AJ....152...97W}, which coupled Bayesian parameter estimation for astrometry with a Gaussian process, to apply to a binary system in order to account for the spatial noise covariance caused by interpolation during data reduction and forming data cubes. Since the squared exponential covariance function would produce improbably smooth noise realizations, we parameterized the spatial noise covariance by the  Mat\'ern ($\nu = 3/2$) covariance function 

\begin{equation}\label{equ:matern}
C_{\ell,ij} = \sigma_i \sigma_j \bigg(1 + \frac{\sqrt{3}r_{ij}}{\ell}\bigg) \text{exp}\bigg(\frac{-\sqrt{3}r_{ij}}{\ell}\bigg)
\end{equation}

\noindent where $\ell$ is the spatial correlation length of noise, $r_{ij}$ is the Euclidean distance between $i,j$ spaxels, and $\sigma_{i}$ is the standard deviation associated with spaxel $i$. The spatial correlation length $\ell$ represents the strength of correlation between two spaxels averaged across the entire wavelength slice.

We remain agnostic to spectral correlation when each wavelength slice is treated independently from one another. However, imposing Gaussian priors on $\theta$ and $\rho$ during the second step does introduce correlation between all wavelength slices. Interpolation and finite spectral resolution also contribute to spectral correlation. The characterization and treatment of spectral correlation is discussed in Section \ref{ssec:atmo}. 

\begin{figure*}[h]
\centering
\includegraphics[scale=.45]{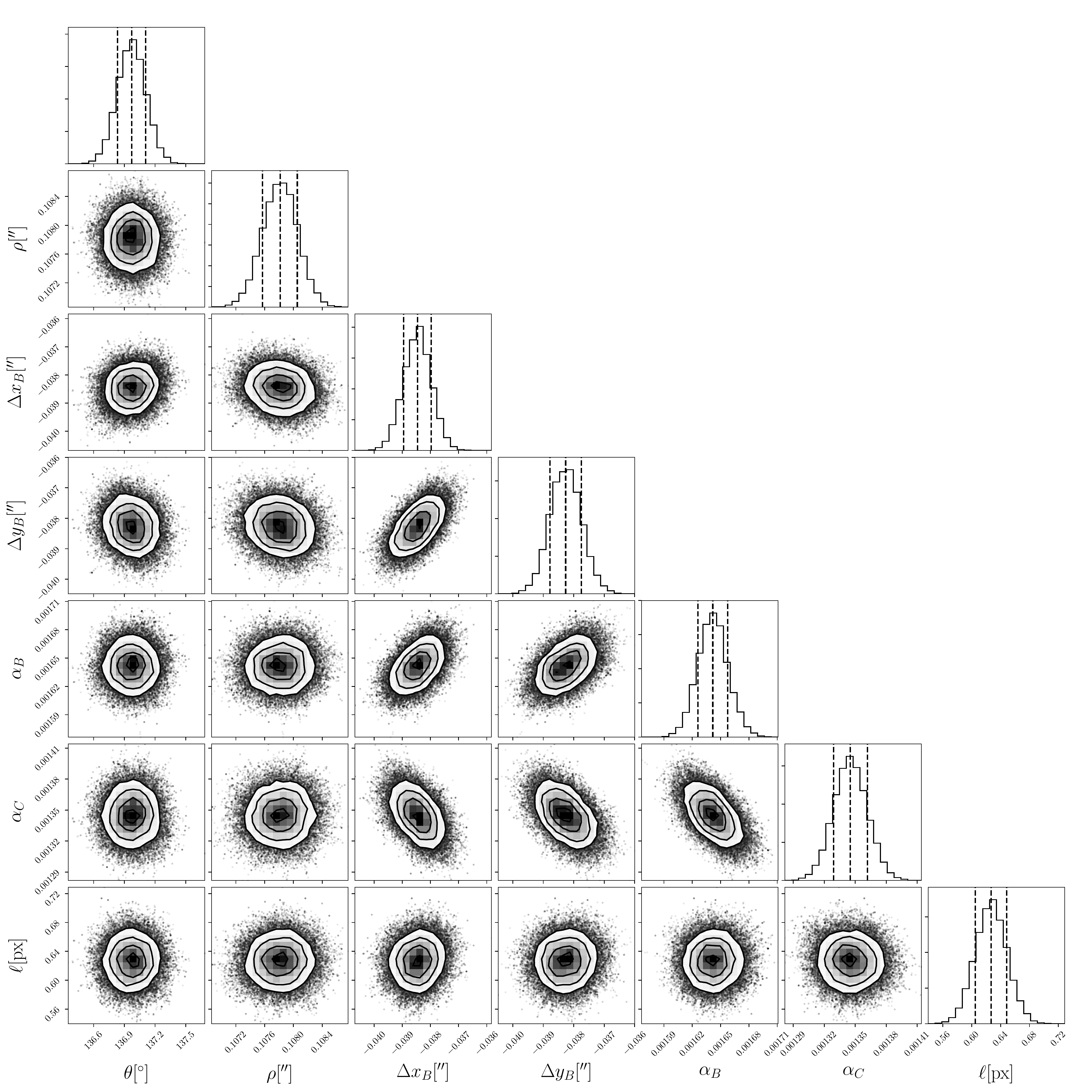} \label{fig:model}
\caption{Posterior distributions for the position of HD~130948B and flux ratios of HD~130948BC with respect to HD~130948A in the representative $3.47 \mu m$ wavelength slice. The posteriors are calculated at each wavelength, and the flux ratios are used to derive the spectrum of HD~130948BC by multiplying with the spectrum of HD~130948A. The contour levels are set to intervals of 0.5-$\sigma$.}
\end{figure*}

The details of the PSF fitting procedure are available in Appendix \ref{sec:fitting}. We evaluated convergence with acceptance rates and the multivariate Gelman-Rubin convergence diagnostic \citep{1992StaSc...7..457G, brooks1998general}. The best fit model at each wavelength was identified as the median of the marginalized posteriors, with 68\%--credible regions of the marginalized posteriors as their uncertainty. Representative best fit models and residuals at the 3.03, 3.47, and 3.87 $\mu m$ wavelength slices of the data cube are shown in Figure \ref{fig:bestfit}, and the marginalized posteriors following the second PSF fitting step for the 3.47 $\mu$m wavelength slice is shown in Figure \ref{fig:model}. The medians of the marginalized posterior distributions from the second step remained within the corresponding 68\%--credible regions of the marginalized posteriors from the first step, with the exception of the far red end (>4 $\mu$m) of the band, where the PSF is larger and the binary is less resolved.

The spatial correlation lengths $\ell(\lambda)$ were determined empirically to trend linearly with wavelength with the form $\ell(\lambda) = (0.115 \pm 0.008 \frac{\text{px}}{\mu \text{m}}) \lambda + (0.236 \pm 0.006 \,\text{px})$. The amplitude of the spatial correlation is not consistent with the diffraction limit of the telescope ($\sim$ 4 px), suggesting interpolation from data cube construction and PSF fitting is contributing to small scale correlation, the ALES data cubes of HD~130948BC are not speckle limited, and the mean photon noise is background dominated. The PSF fitting procedure was repeated for data cubes and registration performed by strictly linear, cubic and quintic interpolation, and permutations thereof, and the spatial correlation lengths remained between 0.5 px and 0.9 px. The Bayesian parameter estimation of the other model parameters were unchanged with the different interpolation schemes. 

The contrast spectrum for each brown dwarf is derived with respect to the primary star HD~130948, a G2V star with solar metallicity \citep{2005ApJS..159..141V}. A model G2V spectrum \citep{1999ApJ...512..377H} was smoothed to the spectral resolution of ALES, and scaled by the WISE W1 (3.3526 $\mu$m; \citealt{2010AJ....140.1868W}) photometry data point of HD~130948A of 5.17 $\pm$ 0.40 Jy \citep{2013A&A...555A..11E}. HD~130948BC do not contribute significantly to the WISE W1 photometry of HD~130948A, as they are 7 magnitudes dimmer. The spatially-resolved contrast spectra of HD~130948B and C were then multiplied by the scaled model to get the absolute flux calibrated spectra for HD~130948BC with uncertainties propagated.

\section{Analysis} \label{sec:analysis}

\subsection{Physical Parameters from Atmospheric Models} \label{ssec:atmo}

\begin{figure*}
\centering
\includegraphics[width=.49\linewidth]{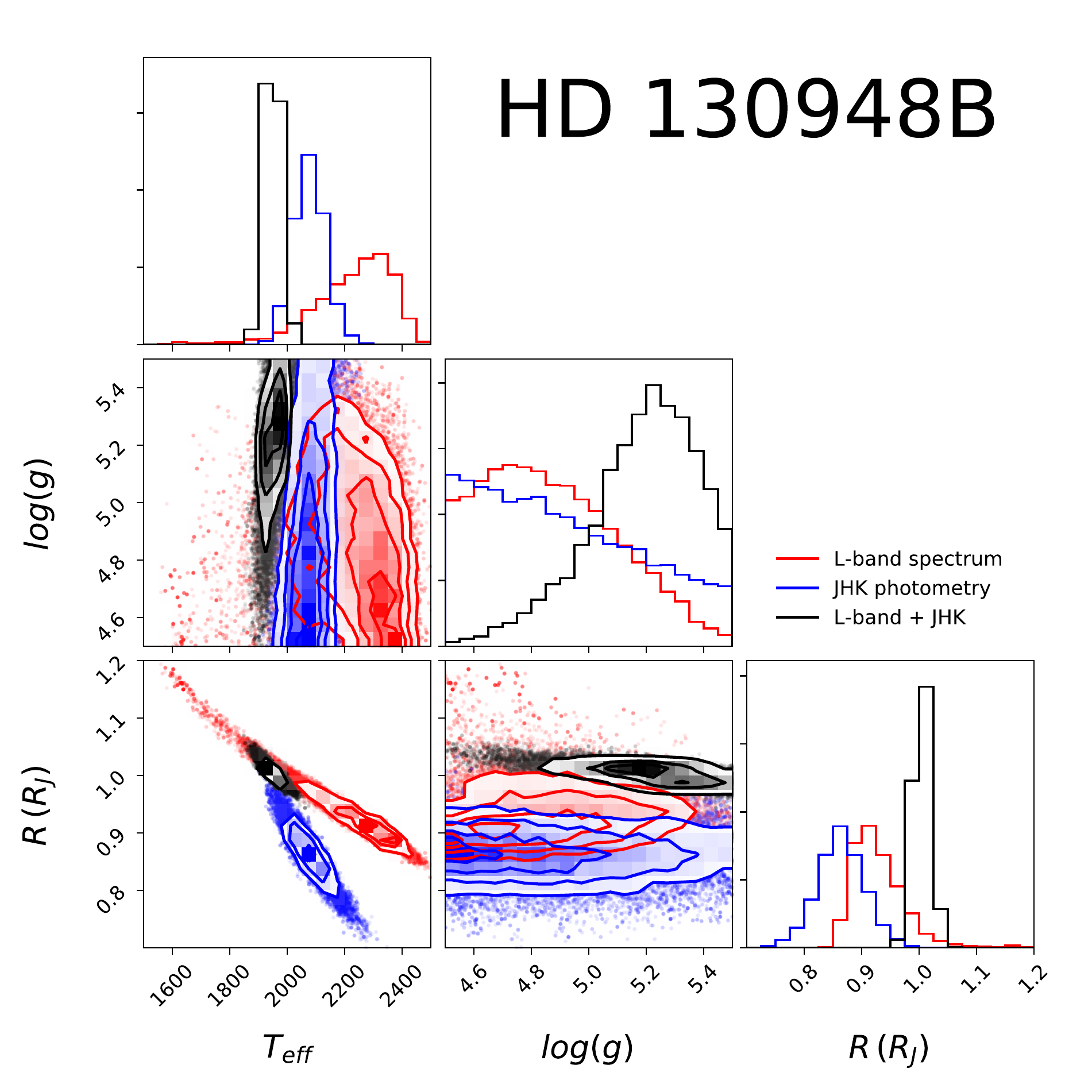}
\includegraphics[width=.49\linewidth]{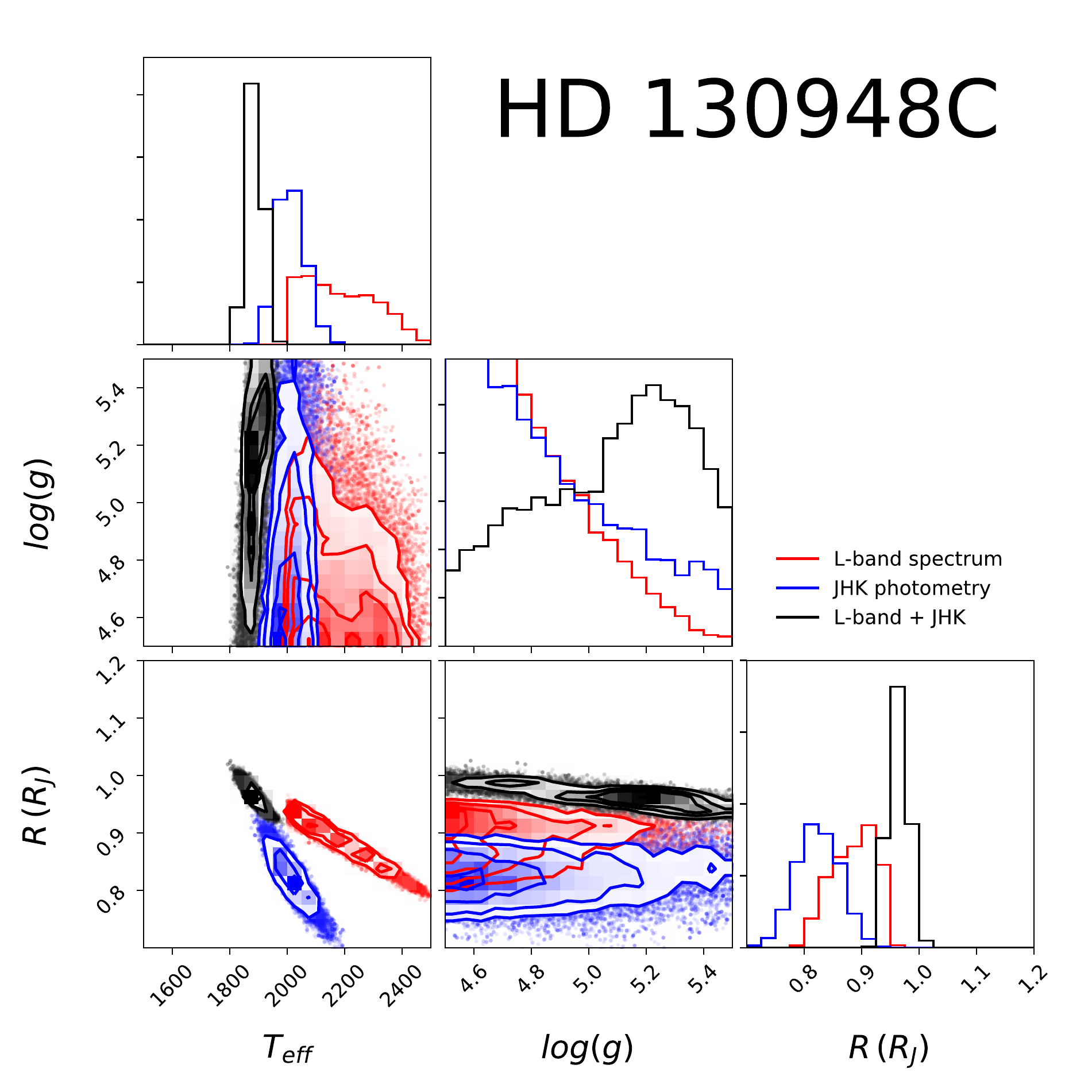}
\caption{Posterior distributions for the near-infrared photometry fits in blue, $L$-band spectral fits in red and near-infrared photometry + $L$-band spectral fits in black for the atmosphere model derived quantities of HD~130948B and C. The contour levels are set to intervals of 0.5-$\sigma$. The medians and credible regions of these quantities are available in Tables \ref{table:B} and \ref{table:c}, respectively. Note that these quantities are heavily correlated; the position of the joint marginalized posteriors not being intermediate between the individual marginalized posteriors is a projection effect, and a result of $JHK$ photometry having limited sensitivity to surface gravity, while $L$-band spectra have limited sensitivity to temperature for hot substellar atmospheres.}
\label{fig:evofits}
\end{figure*}

With the goal of providing a consistent description of the atmospheres of HD~130948BC using $JHK$ photometry and $L$-band spectra, we will explore atmospheric model fitting of solely $L$-band spectra, solely NIR photometry, and joint NIR photometry plus $L$-band spectra.

The PHOENIX atmospheric code outlined in \citet{2011ApJ...733...65B} was used to calculate the synthetic model spectra for this analysis. We chose to interpolate over a pre-synthesized library of model spectra calculated between 1500K and 2500K and log $g$ = 4.5 and log $g$ = 5.5, with resolutions of $\Delta T_{\mathrm{eff}} = 100K,\, \Delta \text{log\,} g = 0.5$ dex. The grid of spectra were log-linear interpolated at the native resolution of the spectra to obtain flux density for arbitrary $k = [T_{\mathrm{eff}},\, \text{log\,}g]$. More sophisticated methods of spectral interpolation, such as Starfish (\citealt{2015ApJ...812..128C}), were not used because the spectral resolution of ALES cubes requires convolution and downsampling that washes out the noding phenomenon described therein.

\subsubsection{Modelling of ALES $L$-band spectra} \label{ssec:fitwithL}

Most integral field spectograph datasets are prone to spectral and spatial correlation. The ALES HD~130948 dataset is no different. In Section \ref{sec:photometry}, we showed there is a non-negligible spatial correlation on subpixel scale. In \citet{2016ApJ...833..134G}, a procedure for modelling spectral correlation is outlined, in which a three component model is proposed to characterize contributions of speckle noise, correlation induced by interpolation during reduction, and uncorrelated noise. With the formalism from Greco \& Brandt, we estimated the correlation $\psi_{ij}$ between pixel values at wavelengths $\lambda_i$ and $\lambda_j$ within a annulus of width 1.5 $\lambda_c$/D in the binary data with the binary masked out ($\lambda_c = 3.50 \mu m$). The correlation was fit with a three-component model, comprising an uncorrelated term with amplitude $A_{\delta}$, spatially-independent Gaussian term with amplitude $A_{\lambda}$ and correlation length $\sigma_{\lambda}$, and a spatially-dependent Gaussian term with amplitude $A_{\rho}$ and correlation length $\sigma_{\rho}$. The respective amplitudes were fit such that $A_{\delta} = 0.55$, $A_{\lambda} = 0.27$, and $A_{\rho} = 0.18$. The noise components were characterized by correlation lengths $\sigma_{\rho} = 0.81$ and $\sigma_{\lambda} = 0.015$. The spectra are correlated to $\sim2$ channels, supporting our assumption we have $N_{\lambda} /2$ resolution elements for critical Nyquist sampling in Appendix \ref{sec:fitting}. This method is not sensitive to the subpixel spatial correlation measured in Section \ref{sec:photometry} because there are no appropriate annuli to measure these lengths.  

Bayes' Theorem is used to write the posterior probability for $k = [T_{\mathrm{eff}},\, \text{log}(g)]$ given observed spectrum $f$ as, $\mathcal{P}(k \,|\, f) \propto \mathcal{L}(f \,|\, k)\, \mathcal{P}(k)$. To quantify the probability of the data conditioned on the model, $\mathcal{L}(f \,|\, k)$, we adopted the following multivariate Gaussian likelihood function (ignoring constants).

\begin{equation}\label{eq:like}
-2\,\text{ln}\,\mathcal{L}(f \,|\, k)= (f \, - \, \alpha F_k)^T\,\Sigma^{-1}\,(f \, - \, \alpha F_k)
\end{equation}

\noindent where $F_k$ is the synthetic spectrum for model $k$, $\Sigma_{ij} = \sigma_i \sigma_j \psi_{ij}$ for flux errors $\sigma$, and $\alpha = (R/D)^2$ for brown dwarf radius $R$ and distance $D$ to the system.

The prior on surface gravity and temperature are defined by the domain limits: $\mathcal{P}(\text{log}(g)[cgs]) = \mathcal{U}[4.5, 5.5]$,  $\mathcal{P}(T_{\mathrm{eff}} [K]) = \mathcal{U}[1500, 2500]$. Realistic priors from evolutionary models could be applicable here. However, \citet{2010ApJ...721.1725D} discuss the existence of a systematic discrepancy between atmosphere and evolutionary model-derived quantities and evolutionary model-predicted quantities from data. It is for this reason that we avoid using the evolutionary models as realistic priors in the atmospheric model fitting suggested in \citet{2016ApJ...833..134G}. 

\textit{Gaia} DR2 provides revised parallax measurements of HD~130948A of $\pi = 54.91 \pm .07$ mas \citep[\textit{Gaia};][]{2016A&A...595A...2G, 2018arXiv180409365G}. The presence of the brown dwarf companions will cause deviations from parallactic trajectory with linear proper motion due to stellar reflex motion, but the lower bound on the orbital period is 155 yr \citep{2013MNRAS.434..671G}; only a fraction of the orbit is traversed during the \textit{Gaia} baseline, suppressing the reflex motion signal. Moreover, HD 130948 is a bright star (G = 5.715) in the \textit{Gaia} catalog, in which stars with G $\lesssim$ 6 have weaker positional accuracy due to saturation of the detector, placing reflex motion signal below the noise floor associated with the parallax uncertainty. We proposed Gaussian priors for parallax $\pi$ to estimate the prior on distance as follows, $\mathcal{P}(D [pc]) = \lvert\frac{\partial \pi}{\partial D}\rvert \cdot \mathcal{N}[\pi; \pi,\sigma_{\pi}^2] = D^{-2}\mathcal{N}[1/D; \pi,\sigma_{\pi}^2]$. The prior for brown dwarf radius was set to $\mathcal{P}(R [R_J]) = \mathcal{LU}[0.5, 1.5]$, with physically motivated limits from evolutionary models and a lack of high insolation.

We used the \citep{2010CAMCS...5...65G} affine-invariant MCMC sampler implemented in the \texttt{emcee} Python package \citep{2013PASP..125..306F} to sample the posterior distribution of $[T_{\mathrm{eff}},\, \text{log\,}g, R, D]$ for HD~130948B and HD~130948C independently. We initialized 50 walkers with a guess $k$ vector plus Gaussian noise at an amplitude of $k \times 10^{-4}$. Each MCMC sampler was run for 400 steps after 400 burn-in steps. We evaluated convergence of the chains with the Gelman-Rubin convergence diagnostic. The resulting marginalized posteriors can be seen in red in Figure \ref{fig:evofits}. Median and credible regions of these posteriors are reported in Tables \ref{table:B} and \ref{table:c}. The best fit $L$-band spectra of HD~130948B and HD~130948C have $\chi_B^2 = 13.7$ and $\chi_C^2 = 21.1$. For 31 spectral bins and free parameters of $k = [T_{\mathrm{eff}},\, \text{log\,}g, \,R$], these appear to be anomalously low $\chi^2$ values. However, the spectral correlation length of $\sim 2$ channels reduces the effective number of degrees of freedom by roughly half.

\subsubsection{Modelling of Near-infrared Photometry}
\citet{2009ApJ...692..729D} measured the MKO $JHK$ photometry of HD~130948BC, which was updated with progressively homogenized analyses in \citet{2014ApJ...790..133D} and \citet{2017ApJS..231...15D}. \citet{2014A&A...566A.130C} presents 2MASS $JHK_s$ photometry that is consistent with the MKO photometry for both brown dwarfs. We chose to use the MKO photometry due to the smaller uncertainty in the photometry. We did not include any covariance in the $JHK$ photometry.

We approach atmospheric modelling of near-infrared photometry using $\chi^2$-fitting. Synthetic photometry $F_k$ was calculated for each model $k$ using MKO filter curves \citep{2002PASP..114..180T}. We used identical priors from Section \ref{ssec:fitwithL}, with the same motivations. To quantify the probability of the data conditioned on the model, we used Equation \ref{eq:like} in the regime where $\Sigma$ is diagonal. We sampled the posterior distribution of $[T_{\mathrm{eff}},\, \text{log\,}g, R, D]$ for HD~130948B and HD~130948C independently with \texttt{emcee} under the same conditions as Section \ref{ssec:fitwithL}, satisfying the same convergence testing, and the resulting marginalized posteriors can be seen in blue in Figure \ref{fig:evofits}. Median and credible regions of these posteriors are reported in Tables \ref{table:B} and \ref{table:c}. The best fit $JHK$ photometry of HD~130948B and HD~130948C have $\chi_B^2 = 6.4$ and $\chi_C^2 = 10.6$.

\begin{figure}[h]

\centering
\includegraphics[trim= 0cm 0cm 0cm 0cm, clip=False, scale=1.1]{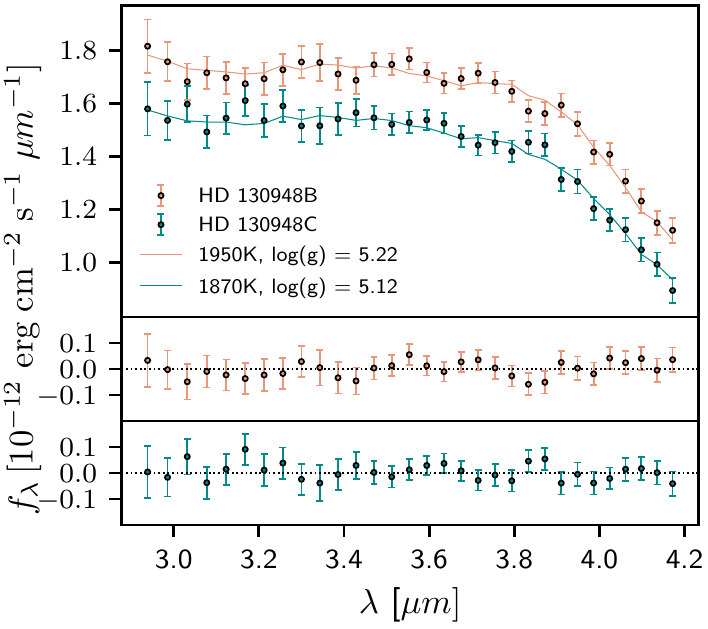}
\label{fig:spectrum}
\caption{Spectra of HD~130948BC with best fit models from \citet{2011ApJ...733...65B} using the near-infrared plus $L$-band fit. The residuals for each fit are plotted below with the color corresponding to the same object in the spectrum.}
\end{figure}

\subsubsection{Modelling of Photometry and Spectra} \label{sssec:photspec}

Combining photometry and spectroscopy to back out meaningful atmospheric properties can unintentionally be driven by weighting schemes. The goodness-of-fit statistic $G_k$ from \citet{2008ApJ...678.1372C} is a commonly used statistic with weights proportional to the wavelength interval associated with the data points. While the $G_k$ allows for heteroskedasticity, it implicitly assumes no correlation between spectral channels.

We approached this problem with two methods: apply no objective weighting scheme and use the covariance measured in $L$-band or apply objective weights according to \citet{2008ApJ...678.1372C} and ignore the covariance in $L$-band. We chose to avoid extending the definition of the goodness-of-fit statistic $G_k$ to include covariance because it is not obvious whether the weighting scheme applies to off-diagonal covariance terms when correlation $\psi_{ij}$ is measured in this manner.

We determined empirically that the posterior derived when applying no objective weights and using the covariance matrix completely contained the posterior when applying objective weights and ignoring the covariance. Therefore, we chose to ignore weights and favor including the covariance matrix. In principle, this means that near-infrared photometry are contributing less to the likelihood function. 

We used identical priors from Section \ref{ssec:fitwithL}, with the same motivations. We also used the same functional form of the likelihood (Equation \ref{eq:like}), with $\psi_{i \neq j} = 0$ for the three photometry points. We sampled the posterior distribution of $[T_{\mathrm{eff}},\, \text{log\,}g, R, D]$ for HD~130948B and HD~130948C independently with \texttt{emcee} under the same conditions as Section \ref{ssec:fitwithL}, satisfying the same convergence testing, and the resulting marginalized posteriors can be seen in black in Figure \ref{fig:evofits}. Median and credible regions of these posteriors are reported in Tables \ref{table:B} and \ref{table:c}. The best fit $JHK$ photometry and $L$-band spectra of HD~130948B and HD~130948C have $\chi_B^2 = 20.1$ and $\chi_C^2 = 31.8$.

Composite spectral energy distributions (SED) of B and C were built using $JHK$ photometry data over their respective bandpasses, the $L$-band spectra of B and C, and filling in the rest of the SED with the best fit model spectra for each suite of fitting procedures \citep[e.g.,][]{2015ApJ...815..108M}. The bolometric luminosities were calculated by integrating these composite spectra. The uncertainty in the bolometric luminosities is derived using a Monte Carlo simulation, taking composite spectra drawn from both a multivariate Gaussian with the mean set to the spectrophotometry and covariance set to $\Sigma$ and model spectra from the respective posteriors, and estimating the standard deviation of their bolometric luminosities. We also performed a Monte Carlo simulation with spectral energy distributions associated with the best fit model spectra of B and C, excluding $JHKL$ data. The median remained unchanged, and the composite spectra method resulted in larger uncertainty. We report the composite SED bolometric luminosity uncertainty in Tables \ref{table:B} and \ref{table:c}.

Figure \ref{fig:spectrum} contains the ALES $L$-band spectra of HD~130948B and C, along with the model fits with parameters set to the medians of the marginalized posteriors from spectral fits for $JHK + L$ spectrophotometry. All following analysis is performed using these median and credible regions, reported in column $JHK + L$ in Tables \ref{table:B} and \ref{table:c}.

\begin{table}
\caption{Atmosphere Model Inferred Properties of HD~130948B}\label{table:B}
\centering
\begin{tabular}{l c c c c c}
\hline\hline
Property & & JHK & L & JHK+L \\[0.5ex]
\hline
$T_{\mathrm{eff}}$ [K]& & $2060 \pm 60$& $2260_{-150}^{+100}$ & $1950 \pm 30$ \\ 
Radius [$R_J$]& & $0.87 \pm 0.04$& $0.92_{-0.03}^{+0.04}$& $1.00 \pm 0.01$ \\ 
log(g) [cm $s^{-2}$]  & & $4.5_{-0.5}^{+0.7}$& $4.7_{-0.4}^{+0.3}$& $5.2 \pm 0.3$\\ 
log($L_{bol}$) [$L_{\odot}$] & & $-3.88 \pm 0.02$ & $-3.87\pm 0.02$&$-3.87\pm 0.01$\\ 
\hline

\end{tabular}

\end{table}

\begin{table}
\caption{Atmosphere Model Inferred Properties of HD~130948C}\label{table:c}
\centering
\begin{tabular}{l c c c c c}
\hline\hline
Property & & JHK & L & JHK+L \\[0.5ex]
\hline
$T_{\mathrm{eff}}$ [K]& & $2000 \pm 60$& $2200_{-180}^{+150}$ & $1870 \pm 30$\\ 
Radius [$R_J$]& & $0.83 \pm 0.04$& $0.87_{-0.03}^{+0.06}$& $0.98 \pm 0.02$\\ 
log(g) [cm $s^{-2}$]  & & $4.4_{-0.4}^{+0.5}$& $4.3_{-0.3}^{+0.4}$& $5.1 \pm 0.3 $\\ 
log($L_{bol}$) [$L_{\odot}$] & & $-3.97 \pm 0.02$ & $-3.97 \pm 0.02$&$-3.96 \pm 0.01$\\ 
\hline

\end{tabular}

\end{table}

\subsection{Individual Masses of HD~130948B and C} \label{ssec:physevo}

Benchmark systems like HD~130948BC provide the rare laboratory necessary to obtain individual masses of brown dwarfs, a measurement that is crucial to tests of evolutionary models. For most directly imaged planets, we must rely on evolutionary models to convert luminosity to mass given an age estimate (and potentially information on formation and initial entropy), so putting such models to the test is essential. In order to do this test, we need to isolate the mass of each object using the total mass constraint from orbital monitoring and our measurements of bolometric luminosity.

Substellar objects will tend to radiatively cool with time \citep{1991ARA&A..29..163S, 1993RvMP...65..301B}; evolutionary models of substellar objects propose a luminosity-age-mass relationship, owing to a lack of a sustainable source of internal energy from nuclear reactions, with more massive objects starting hotter and more luminous. We exploit this relationship with evolutionary models from \citealt{2015AA...577A..42B} and \citealt{2008ApJ...689.1327S} to isolate the masses and age of HD~130948BC. Since the measured luminosities are nearly equal, extreme mass ratios can be ruled out. Therefore, breaking the mass degeneracy is primarily driven by the tight dynamical mass and bolometric luminosity constraints instead of the evolutionary models themselves. 

\begin{table}
\caption{Evolutionary Model Inferred Properties of HD~130948BC}
\centering
\begin{tabular}{l c c c c}
\hline
\hline
Property & & HD~130948B & HD~130948C  \\[0.5ex]
\hline

\multicolumn{4}{c}{Input Observed Properties}\\
\hline
Mass [$M_{J}$] &&\multicolumn{2}{c}{$116.2_{-0.8}^{+0.9}$}   \\ 
log($L_{bol}$) [$L_{\odot}$] & & $-3.87\pm0.01$& $-3.96\pm0.01$\\
\hline
\multicolumn{4}{c}{\citealt{2008ApJ...689.1327S} Hybrid Models}\\

\hline
Mass [$M_{J}$] & & $59.8\pm0.6$& $56.4\pm0.6$  \\ 
log($L_{bol}$) [$L_{\odot}$] & & $-3.87\pm0.01$ & $-3.96\pm0.01$ \\ 
q [$M_C / M_B$] & & \multicolumn{2}{c}{$0.94 \pm 0.01$} \\ 
Age [Gyr] & & \multicolumn{2}{c}{$0.45\pm 0.01$}  \\ 
$T_{\mathrm{eff}}$ [K]& & $1900\pm 20$& $1800\pm20$\\ 
Radius [$R_J$]& & $1.037\pm.002$ & $1.037\pm.002$  \\ 
log(g) [cm $s^{-2}$] & & $5.14\pm 0.01$& $5.11 \pm 0.01$ \\ 
\hline
\multicolumn{4}{c}{\citealt{2015AA...577A..42B} Models}\\

\hline
Mass [$M_{J}$] & & $59.8\pm0.6$& $56.3\pm0.5$\\ 
log($L_{bol}$) [$L_{\odot}$] & & $-3.86\pm 0.01$ & $-3.97\pm 0.01$  \\ 
q [$M_C / M_B$] & & \multicolumn{2}{c}{$0.94\pm0.01$} \\ 
Age [Gyr] & & \multicolumn{2}{c}{$0.51_{-0.02}^{+0.01}$} \\ 
$T_{\mathrm{eff}}$ [K]& & $1960 \pm 20$& $1840\pm 20$ \\ 
Radius [$R_J$]& & $0.991_{-0.002}^{+0.005}$& $0.990_{-0.002}^{+0.005}$ \\ 
log(g) [cm $s^{-2}$]  & & $5.18 \pm 0.01 $& $5.16 \pm 0.01$\\ 
\hline

\end{tabular}
\label{table:evolution}
\end{table}

For individual ages between 1 Myr and 1 Gyr, we calculate two masses with the prior that their sum is distributed according to the dynamical mass posterior. With revised parallax from \textit{Gaia} DR2, the total dynamical mass of HD~130948BC derived from astrometric analysis reported in \citep{2017ApJS..231...15D} of $M_{dyn} = 115.4^{+2.2}_{-2.1}\,M_J$ is updated to $M_{dyn} = 116.2^{+0.9}_{-0.8}\,M_J$. The model bolometric luminosity is calculated by log-linearly interpolating the evolutionary model grids. The log-likelihood is calculated from the residuals of model luminosities and measured luminosities of HD~130948BC and their respective errors. Uniform bounded priors were used for the age of the system. The posterior distribution of masses of each component and the age of the system was sampled using \texttt{emcee}. The credible regions from the resulting marginalized posteriors are reported for both sets of models in Table \ref{table:evolution}. The apparent bolometric fluxes from HD~130948B and C at best fit are $(1.31 \pm 0.03) \times 10^{-11} \mathrm{erg \,s^{-1}\, cm^{-2}}$ and $(1.07 \pm 0.03) \times 10^{-11} \mathrm{erg \,s^{-1} \,cm^{-2}}$, respectively.

\section{Discussion} \label{sec:discussion}

\begin{figure*}
\centering
\includegraphics[trim=0cm 0cm 0cm 0cm,clip=False, width=\textwidth]{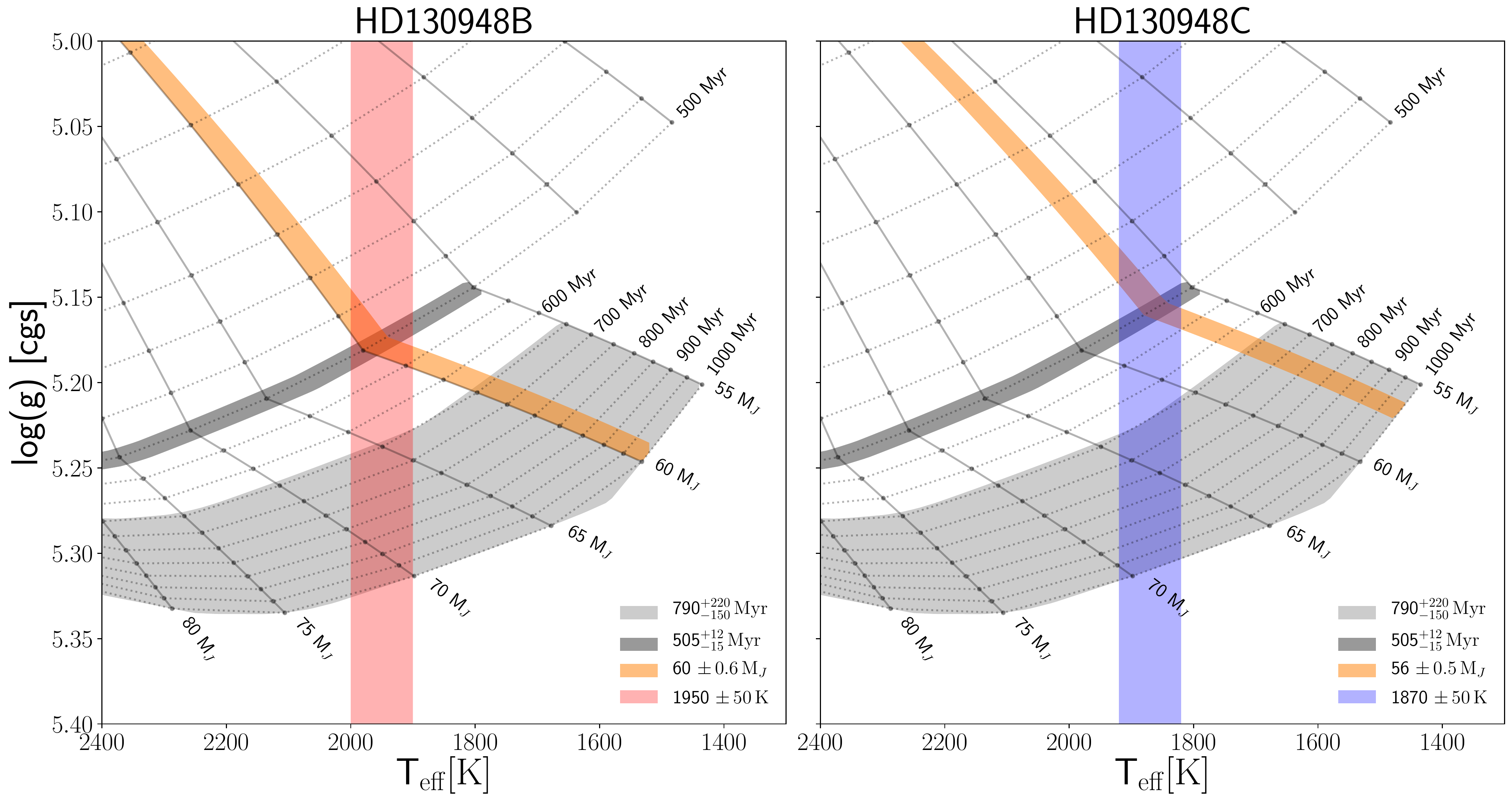} \label{fig:gridthing}
\caption{Isomass lines in solid black and isochrones in dashed black lines from \citealt{2015AA...577A..42B}. The light gray region corresponds to the gyrochronology constraint from the primary star. The dark gray region corresponds to the 68\%--credible regions of age derived from the evolutionary model fitting of HD~130948BC according to Section \ref{ssec:physevo}. The orange regions correspond to the credible regions of mass derived from the evolutionary model fitting of HD~130948BC. The red and blue regions correspond to the credible regions of temperature derived from the atmospheric model fitting of HD~130948BC.}
\end{figure*}

Independent validation of atmospheric and evolutionary models is critical for characterizing the atmospheres of directly imaged extrasolar planets and brown dwarfs. The masses of directly imaged exoplanets and brown dwarfs are generally poorly constrained owing to the difficulty in independently measuring masses, and inferences being extremely model-dependent. Yet mass is fundamentally important to test models of giant planet formation and empirically calibrate substellar evolutionary models. HD~130948B and C are some of the only brown dwarfs with independently measured mass, age, and luminosity to probe these models \citep{2009ApJ...692..729D, 2010ApJ...711.1087K, 2014ApJ...790..133D, 2017ApJS..231...15D}. 

\subsection{Evolutionary and Atmospheric Models}
The atmosphere model fitting is independent of the evolutionary models (apart from physically motivated bounds on the radius prior). The evolutionary model fits are only informed by the modeled bolometric luminosities of HD~130948B and C, which is driven by the data, and independent dynamical mass measurements. The different models are not strictly expected to derive consistent atmosphere quantities. Specifically, near-infrared spectroscopy has been shown to produce atmospheric model fits that are discrepant by 250K from evolutionary model fits to the same data set \citep{2010ApJ...721.1725D}.

Our fits to $L$-band spectra have similarly derived temperatures $\sim 250$ K warmer than the other two methods (Tables \ref{table:B} and \ref{table:c}). The coarse sampling of atmospheric parameters for the atmospheric model grid also limits what can be determined from the surface gravity measurement. At ALES spectral resolution and this temperature regime, $L$-band does not vary significantly with surface gravity, and there are no prominent features that are highly gravity dependent. While $L$-band is not particularly diagnostic of atmospheric parameters for hotter objects, $L$-band becomes critical at lower temperatures \citep[e.g.,][]{2014ApJ...792...17S, 2015ApJ...804...61B}.

However, atmospheric model fitting to the combination of $JHK$ photometry and $L$-band spectra for HD~130948B and C has resulted in posteriors that are completely consistent with evolutionary model-derived quantities, illustrating the importance of extended wavelength coverage for substellar objects. Figure \ref{fig:gridthing} depicts isomass lines and isochrones in effective temperature-surface gravity space with evolutionary models from \citet{2015AA...577A..42B}. The red and blue regions denote the credible regions of effective temperature (and surface gravity) from atmosphere model fitting, including the grid spacing errors set to half the grid spacings ($\sigma_{T_{\mathrm{eff}}} = 50K,\, \sigma_{ \text{log\,} g} = 0.25$ dex). The publicly available evolutionary models did not continue past 1 Gyr in this mass regime. 

\subsection{Age of HD~130948}
We assume coevolution of the binary brown dwarfs with the primary star for the following discussion. With the method described in Section \ref{ssec:physevo}, the Baraffe et al. 2015 (BHAC15) and hybrid Saumon \& Marley 2008 (SM08) evolutionary models were used to derive an age of $0.51_{-0.02}^{+0.01}$ Gyr and $0.45\pm0.01$ Gyr for HD~130948BC, respectively. Both evolutionary model-derived ages are consistent with the age of HD~130948A as traced by the Ca II HK emission of $0.5 \pm 0.3$ Gyr \citep{2009ApJ...692..729D}, the previous evolutionary model-derived age of $0.44 \pm 0.04$ \citep{2017ApJS..231...15D}, and the \citet{2007ApJ...669.1167B} relationship for gyrochronological age of $0.65_{-0.10}^{+0.13}$ Gyr. 

The gyrochronology relation from \citet{2008ApJ...687.1264M} results in an age of $0.79_{-0.15}^{+0.22}$ Gyr does remain an outstanding topic of discussion, due to its adoption as the age of the system in \citet{2009ApJ...692..729D}. 

One factor in this adoption is the observation that the $B-V$ color of HD~130948A suggests an age marginally consistent with, if not older than, the Hyades cluster. The Hyades cluster was believed to have a tight age constraint of 625 $\pm$ 50 Myr \citep{1998A&A...331...81P}. However,  \citet{2015ApJ...807...58B, 2015ApJ...807...24B} have fit rotating stellar models to main-sequence turnoff Hyads to measure the age of the Hyades cluster to be older and with wider spread (750 $\pm$ 100 Myr). If the Hyades are systematically older, gyrochronology relations would need to be re-calibrated to ameliorate the updated age \citep{2016ApJ...822...47D}. It should also be noted that \citet{2018ApJ...863...67G} used a different prescription of rotating stellar models and derived an age $\sim 680$ Myr, which is roughly consistent with the canonical age of the cluster. For HD~130948A to be strictly older than the Hyades, the bolometric luminosities of HD~130948B and C would be considerably overluminous compared to the predictions from evolutionary models.

Several possible explanations exist for the age discrepancy, including (1) the treatment of clouds, metallicities, atmospheric opacities plays a major role evolutionary models \citep{2011ApJ...736...47B}, (2) evolutionary models systematically overpredict cooling rates of substellar objects \citep{2011ASPC..448..111D}, (3) very strong, interior magnetic fields inhibit the onset of convection in HD~130948BC \citep{2010ApJ...713.1249M}, (4) the efficiency of convection that decreases for fast-rotating, highly magnetic low-mass stars extends to substellar evolution \citep{2007A&A...472L..17C}, (5) atypical stellar rotation can be induced from formation via gravitational instability in a long-lived, massive circumstellar disk \citep{2014ApJ...790..133D}, and (6) systematic offsets in gyrochronology relations for field G stars \citep{2008ApJ...687.1264M}. Probing these explanations is beyond the scope of this paper, but will become critical to investigate cooler objects with any precision.

\begin{figure*}[h]
\centering
\includegraphics[trim= 0cm 0cm 0cm 0cm, clip=False, width=\textwidth]{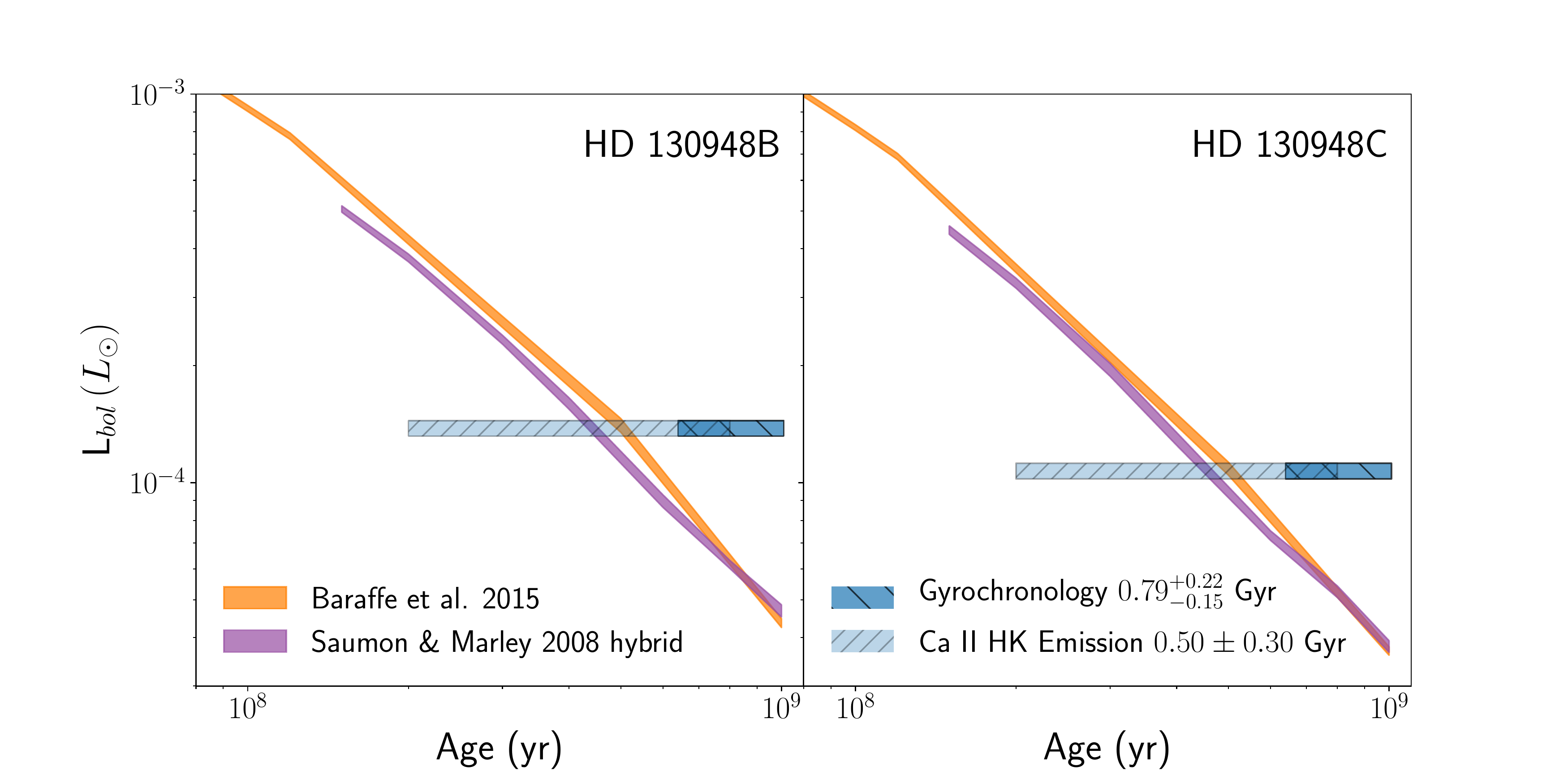} \label{fig:evo}
\caption{Isomass lines from \citealt{2015AA...577A..42B} and \citealt{2008ApJ...689.1327S} evolutionary models for the best fit masses of HD~130948BC with the thickness of the line corresponding to the 68\%--credible region in mass. The two blue boxes correspond to the two age constraints derived in \cite{2009ApJ...692..729D} for the age of the primary HD~130948A through gyrochronology and chromospheric activity.}
\label{fig:fits}
\end{figure*}

Luminosity evolution for the HD~130948B and C are depicted in Figure \ref{fig:evo}. The individual lines and linewidth correspond to the median and 68\%--credible regions of mass of best-fit as derived in Section \ref{ssec:physevo}, propagated from 100 Myr to 1 Gyr using the two evolutionary models. This plot is qualitatively identical to Figure 10. of \citet{2009ApJ...692..729D}. 

The consistency of atmospheric quantities derived from evolutionary model fitting and atmospheric model fitting for HD~130948BC provides complementary evidence to support the age derived from the evolutionary model fits. We chose to focus on BHAC15 evolutionary models. A Monte Carlo simulation was performed using draws of the atmosphere model-derived ($JHK + L$) effective temperature posteriors and the individual mass posteriors derived in the evolutionary model fitting, and propagated through the evolutionary model grid to obtain an average age for each draw of an effective temperature and mass for B and C. The age derived using BHAC15 models was $0.50 \pm 0.07$ Gyr, which is consistent with the age traced by Ca II HK emission and the \citet{2007ApJ...669.1167B} gyrochronology age. 

The derived age of HD~130948BC is younger than age estimates of the Hyades, while the  $B - V$ color and rotational period of HD~130948A lie in parameter space beyond the 625 Myr isochrone of the Hyades \citep{1997A&A...323L..49P, 2000AJ....120.1006G, 2008ApJ...687.1264M}. Leveraging the consistency of atmospheric and evolutionary models for HD~130948BC and assuming the binary and primary are coeval, we invoke anomalous stellar angular momentum loss as an explanation of the systematically older age estimates from gyrochronology relationships \citep{2007ApJ...669.1167B, 2008ApJ...687.1264M} from evolutionary model-derived ages for HD 130948. 

This explanation, however, is not sufficient to describe the behavior of other over-luminous substellar objects, such as Gl 417BC. The Gl 417 system is a hierarchical triple system similar to HD 130948, with the Gl 417BC brown dwarf binary separated 90 arcseconds from their primary star Gl 417A, and therefore ineffective at driving anomalous stellar angular momentum loss \citep{2014ApJ...790..133D}.

\begin{figure*}[h]

\centering
\includegraphics[trim = 0cm 0cm 0cm 0cm, width=\textwidth]{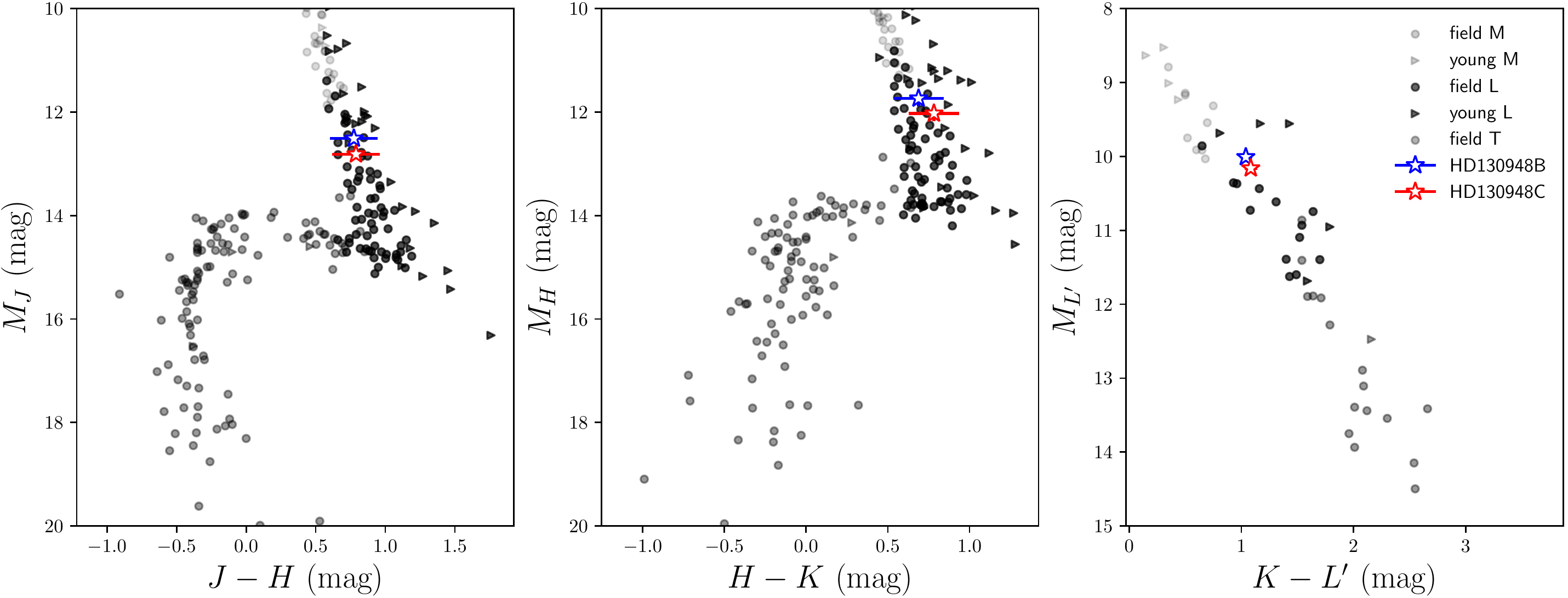}
\label{fig:cmd}
\caption{Color magnitude diagram for field M, L and T brown dwarfs with HD 130498 BC. The data for the field brown dwarfs and the $JHK$ magnitudes of HD 130948BC are from the Database of Ultracool Parallaxes \citep{2012ApJS..201...19D, 2013Sci...341.1492D, 2016ApJ...833...96L} and \citep{2017ApJS..231...15D}, respectively.}
\end{figure*}

\subsection{Spectrophotometric Characterization of HD~130948BC}

Section \ref{ssec:atmo} presents fitting the spectral energy density with three distinct methods, each of which produce broadly consistent atmosphere parameters for two L4 brown dwarfs. Around 1800-2000$K$ brown dwarfs, the spectral features in $L$-band are dominated by the H${}_2$O pseudo-continuum \citep{2015ApJ...804...61B}. At ALES resolution, cooler temperature brown dwarfs (<1800$K$) begin to exhibit CH${}_4$ PQR-branch absorption that suppress the water pseudo-continuum near 3.3 $\mu m$. No significant methane absorption is evident in the $L$-band spectra of HD 130948BC, placing the pair earlier than L5 spectral type. Due to the lack of spectroscopic standards in $L$-band, we defer to the spectral type determination of L4 $\pm$ 1 from the literature \citep{2002ApJ...567L..59G}.

The color magnitude diagram with field brown dwarfs and HD~130948BC is plotted in Figure \ref{fig:cmd}. Synthetic $L^{\prime}$/$W1$ photometry was calculated by using the MKO-$L'$/WISE-$W1$ filter curve and the ALES spectra of HD~130948BC. Synthetic $L^{\prime}$ magnitudes of HD~130948BC are $11.308\pm0.034$ mag and $11.461 \pm 0.039$ mag, respectively. The delta $L^{\prime}$ magnitude between HD~130948B and HD~130948C is $0.153 \pm 0.034$. The calculated $K - L^{\prime}$ for HD~130948BC are $1.040 \pm 0.042$ mag and $1.080 \pm0.042$ mag, respectively. The synthetic $W1$ magnitudes of HD~130948BC are $11.727\pm0.036$ mag and $11.856 \pm 0.043$ mag, respectively. The delta $W1$ magnitude between HD~130948B and HD~130948C is $0.129 \pm 0.037$. The calculated $K - W1$ for HD~130948BC are $0.622 \pm 0.042$ mag and $0.689 \pm0.042$ mag, respectively.

Near-infrared spectra of HD~130948BC exist, but the observation suffered from differential slit loss \citep{2002ApJ...567L..59G}. The continuum contains the temperature and gravity information, so we chose not to include this dataset for spectral fitting. However, Goto et al. did identify 2MASSW J00361617+1821104 (L4) as being the best-matched template spectrum for both the observed HD~130948BC medium resolution $H K$ spectra. Photometry of the 2MASSW J00361617+1821104 was used to calculated $K - L^{\prime}$ of $0.96 \pm 0.058$ \citep{2002ApJ...564..452L, 2004AJ....127.3553K}. Comparing the three L4 brown dwarfs, both HD~130948BC are slightly redder, perhaps due to adolescence.

There also exists resolved optical photometry of the binary from HST/ACS-HRC \citep{2009ApJ...692..729D}. These data comprise flux ratios of HD~130948BC in four red-optical bandpasses. In order to simultaneously fit red-optical photometry, NIR photometry and $L$-band spectroscopy, our model grid would need to be expanded considerably to cover a much broader range of parameters (e.g., abundances, cloud properties, non-equilibrium chemistry). An extensive parameter search is beyond the intended scope of this paper. The synthetic flux ratios calculated for the $JHK+L$ best fit for $F850LP$, $F R914M \,(8626\,\angstrom)$, $F R914M\, (9402 \,\angstrom)$, and $F R914M \,(10248 \,\angstrom)$ are $0.10 \pm 0.03$, $0.31 \pm 0.03$, $0.40 \pm 0.03$, and $ 0.44 \pm 0.02$ mag, respectively.

\section{Conclusion} \label{sec:conclusions}

We obtained 2.9--4.1 micron spectra of HD~130948BC with the ALES integral field spectrograph. This is the first time an adaptive optics-fed integral field spectrograph has been used at these wavelengths. We demonstrated that atmospheric models are able to reproduce the spectral energy distributions of these benchmark brown dwarfs. The $JHK$ photometry and $L$-band spectra become potent constraints when used in tandem, recovering parameters consistent with the evolutionary model fits. Our results suggest low-resolution $L$-band spectra can ameliorate the discrepancy of atmosphere fits and evolutionary model fits for objects with only $JHK$ photometry, making ALES a powerful tool in aiding our understanding of evolutionary and atmosphere models.

Our determination that the ALES spectra can aid near-infrared measurements in characterizing the atmospheres of HD~130498BC has been the culmination of the development of a versatile pipeline and observation strategy that sets ALES as a new instrument capable of characterizing the thermal spectral properties of directly imaged planets and brown dwarfs.  When JWST launches, there will be enormous scientific opportunity for studying exoplanets in the thermal infrared.  ALES will be complementary to JWST.  While ALES is less sensitive, it probes smaller inner working angles, especially in the context of spectroscopy.  Both JWST and ALES will increase the wavelength range over which we study directly-imaged planets, which will be especially important as we begin to study colder exoplanets.

\section{Acknowledgements}

We would like to thank the anonymous referee for the constructive criticism that has greatly benefitted the clarity of our manuscript.

The LBT is an international collaboration among institutions in the United States, Italy and Germany. LBT Corporation partners are: The University of Arizona on behalf of the Arizona university system; Istituto Nazionale di Astrofisica, Italy; LBT Beteiligungsgesellschaft, Germany, representing the Max-Planck Society, the Astrophysical Institute Potsdam, and Heidelberg University; The Ohio State University, and The Research Corporation, on behalf of The University of Notre Dame, University of Minnesota and University of Virginia. This paper is based on work funded by NSF Grants 1608834 and 1614320.

JMS is supported by NASA through Hubble Fellowship grant HST-HF2-51398.001-A awarded by the Space Telescope Science Institute, which is operated by the Association of Universities for Research in Astronomy, Inc., for NASA, under contract NAS4-26555. Barman is supported by NSF AAG awards 1405505 and 1614492 and NASA XRP award NNX17AB63G. ZWB is supported by the National Science Foundation Graduate Research Fellowship under Grant No. 1842400.

\software{Astropy \citep{2018AJ....156..123A}, matplotlib \citep{Hunter:2007}, SciPy \citep{scipy}, NumPy \citep{2011CSE....13b..22V}, emcee \citep{2013PASP..125..306F}, jupyter \url{http://jupyter.org/}, SAOImage DS9}

\facility{LBT (LBTI/LMIRcam/ALES)}

\bibliographystyle{yahapj}
\bibliography{references}

\appendix
\section{Bayesian Parameter Estimation for PSF Fitting} \label{sec:fitting}

In the following procedure, the subscript $\lambda$ will be dropped for clarity; the procedure implicitly applies to each wavelength slice in the data cube. For each wavelength slice of the PSF and binary, there are six parameters $\phi$ defining a model of the two brown dwarfs with two shifted PSFs: the position ($y_B$, $x_B$) of HD~130948B, the position angle $\theta$ and the projected separation of the brown dwarfs $\rho$, the contrast ratios $\alpha_B$ and $\alpha_C$ of HD~130948BC with respect to the PSF, HD~130948A, and a hyperparameter $\ell$ corresponding to a spatial correlation length. An additive factor to quantify sky background offsets was determined empirically to be consistent with zero, and therefore not included.

The binary $\mathcal{D}$ is modeled by shifting and scaling the PSF, $\mathcal{A}$, to each object. The unscaled model for HD~130948B is $\mathcal{A}_B(\phi) = \mathcal{A}[y_B, x_B]$ and the unscaled model for HD~130948C is $\mathcal{A}_C(\phi) = \mathcal{A}[y_B + \rho\, \text{sin}\,\theta,\, x_B + \rho\, \text{cos}\,\theta]$, where $[\cdot,\,\cdot]$ denotes the translation function of an image. The binary was coarsely centered in a $17 \times 17$ pixel fitting region that defines the reference origin. The PSF was centered in a square fitting region of area that was ten pixels larger on all sides to ensure that any plausible shift of the PSF would populate the entire fitting region with data. The residuals between the data and the model is $R \equiv \mathcal{D}\, - \,\alpha_B \mathcal{A}_B(\phi) \, - \,\alpha_C \mathcal{A}_C(\phi)$. 

The variance images of $\mathcal{A}_B$, $\mathcal{A}_C$, and $\mathcal{D}$ are propagated similarly, and are denoted $\sigma^2_B$, $\sigma^2_C$, and $\sigma^2_{\mathcal{D}}$. The uncorrelated uncertainty vector is $\sigma^2(\phi) = \sigma_{\mathcal{D}}^2 + \alpha_B^2 \sigma_B^2+\alpha_C^2\sigma_C^2$, which is converted into the covariance matrix $C_{\ell}$ using Equation \ref{equ:matern}. In principle, variance is not linear and therefore $\sigma^2_B$ and $\sigma^2_C$ cannot be interpolated simply by this translation function. However, the contribution of the PSF variance scaled by the square of the flux ratio already acts as a small perturbation of the variance image of the binary data. At the measured spatial correlation lengths, $\ell$, the contribution of the covariance terms is even higher order and therefore neglected. That being said, $C_{\ell}$ is still highly nonlinear in $\phi$ and therefore this problem is not ideal for generalized least square estimators.

Bayes' Theorem is used to write the posterior probability for $\phi$ and $\ell$ given $\mathcal{D}$ as, $\mathcal{P}(\phi, \ell | \mathcal{D}) \propto \mathcal{L}(\mathcal{D} | \phi, \ell) \mathcal{P}(\phi, \ell)$. To quantify the probability of the data conditioned on the model, we adopted a multivariate Gaussian likelihood function.
\begin{equation}
-2\,\text{ln}\,\mathcal{L}(\mathcal{D} | \phi, \ell) = R^T\,C_{\ell}^{-1}\,R + \text{ln}|C_{\ell}| + N_{\text{pix}} \text{ln} \,2 \pi
\end{equation}
We employ a uniform, bounded priors $\mathcal{P}(\phi, \ell)$. The bounds exclude PSF shifts off the fitting region, negative separations $\rho$ of the PSFs or $\theta \pm \pi$ (corresponding to $y_B$, $x_B$ instead describing the position of HD~130948C), and extremely large or small $\alpha_B$ and $\alpha_C$. Spatial correlation lengths are positive-definite and bounded above by the size of the fitting region. 

We used \texttt{emcee} to sample the posterior distribution for each wavelength slice independently. We initialized 100 walkers with a guess $\phi$ vector plus Gaussian noise at an amplitude of $\phi \times 10^{-4}$. The MCMC sampler was run for 1000 steps after 1000 burn-in steps for each wavelength. The matrix inversion, multiplication and determinant of the Hermitian, positive-definite covariance matrix was calculated using the Cholesky decomposition to take advantage of numerical stability.

The resulting posterior distributions at each wavelength were marginalized over position, contrast and spatial correlation length terms, and combined jointly to calculate the median values for position angle and separation across all wavelengths, denoted $\theta^*$ and $\rho^*$. The plate scale error of 0.1 mas/spaxel was propagated with Monte Carlo when converting the separation in spaxels to angular separation in mas. The standard deviations $\sigma_{\theta^*}$ and $\sigma_{\rho^*}$ were calculated from the joint distribution assuming critical Nyquist sampling. This assumption is supported in Section \ref{ssec:fitwithL}. The values adopted for position angle and separation are $\theta^* = 137.0 \pm 0.2^{\circ}$ and $\rho^* = 107.8 \pm 0.3 $ mas, respectively. 

The second step uses the same uniform, bounded priors $\mathcal{P}(\phi)$ from the first step, except for position angle and separation. The prior for position angle and separation are updated to be $\mathcal{P}(\theta) = \mathcal{N}[\theta; \theta^*,\sigma_{\theta^*}^2]$ and $\mathcal{P}(\rho) = \mathcal{N}[\rho; \rho^*,\sigma_{\rho^*}^2]$. The posterior distribution was sampled again for each wavelength under the same conditions stated before. 

The correlations evident in Figure \ref{fig:model} manifest as a weak degeneracy between position of the brown dwarfs and flux ratios with respect to the PSF: the model binary built from two shifted PSFs will "exchange" flux when they are mislocated. A precision astrometric solution or $L$-band flux ratios of the brown dwarfs would be necessary to break this degeneracy (with more assumptions). Without such information, we do not probe this degeneracy further.

\end{document}